\documentclass[12pt]{article}
\usepackage{amsmath,amssymb,amsthm}

\newtheorem*{definition*}{Definition}
\newtheorem*{theorem*}{Theorem}
\newtheorem*{proposition*}{Proposition}
\newtheorem*{example*}{Example}
\newtheorem*{exercise*}{Exercise}
\newtheorem*{corollary*}{Corollary}
\newtheorem*{remark*}{Remark}

\usepackage{geometry}
\usepackage[shortlabels, inline]{enumitem}
\usepackage{bm}
\usepackage{graphicx}
\usepackage{float}

\allowdisplaybreaks[2]
\usepackage{color}
\usepackage{mathtools}
\mathtoolsset{showonlyrefs=true}
\usepackage{url}

\begin{document}

\begin{center}
  ~\vspace{20pt}
  
  \Large 
  Dynamics of Periodic Bubbles and Crashes: Modeling Market Overheating and Panic Selling via Cubic Momentum

  \vspace{20pt}
  
  \large
  Naohiro Yoshida

  \normalsize
  Department of Economics,
  Keiai University

  1-5-21,
  Anagawa, Inage,
  263-8588,
  Chiba,
  Japan

  E-mail:
  \url{n-yoshida@u-keiai.ac.jp}

\end{center}

  \vspace{20pt}

  \noindent
  MSC 2020:
  91G15    ;
  91G80    ;
  37N40    ;
  60G55

  \noindent
  JEL Classification:
  G01   ;
  G12   ;
  C63

  \noindent
  Keywords:
  Financial bubbles    ;
  Market crashes    ;
  Nonlinear dynamics    ;
  Hawkes process    ;
  Momentum

\begin{abstract}
This paper proposes a simple and parsimonious discrete-time simulation model to describe the endogenous formation and periodic collapse of financial bubbles. While existing literature has extensively explored the statistical properties of locally explosive bubble dynamics, capturing the micro-level interplay of investor herd behavior and panic selling within a unified framework remains a challenge. Our model addresses this by introducing a cubic function of market momentum to determine the balance of trading directions. This mechanism drives both trend-following behavior during the bubble phase and sudden market crashes when the momentum exceeds a critical threshold. Furthermore, inspired by the self-exciting nature of the Hawkes process, the model endogenizes``market frenzy" by linking trading frequency directly to the accumulated momentum. Simulation results demonstrate that this minimal setup successfully replicates the complex, nonlinear dynamics of bubbles, including simultaneous surges in liquidity and price, followed by dramatic crashes.
\end{abstract}

\newpage

\section{Introduction}

Understanding the mechanisms behind the formation and subsequent rapid collapse (crash) of periodically occurring bubbles in financial markets is a critical issue in economics and financial engineering. To date, various mathematical and time-series models have been proposed to describe the dynamics of rational bubbles, wherein market prices deviate from economic fundamentals, e.g., \cite{flood1980market, tirole1985asset, diba1988explosive, evans1991pitfalls, froot1991intrinsic}. 

A prominent early approach is the discrete-time model of stochastic bubbles proposed by \cite{blanchard1982bubbles}. This model captures the states of whether a bubble continues to grow or collapses, offering a more realistic approach compared to deterministic bubbles. 
However, stochastic bubbles suffered from a theoretical limitation: once a bubble completely collapses, it cannot restart (or grow) again. 
To overcome this limitation, \cite{fukuta1998simple} introduced the concept of ``incompletely bursting bubbles." 
By constructing a simple and tractable discrete-time model in which a bubble does not drop to absolute zero, \cite{fukuta1998simple} demonstrated that bubbles can resume growth following an incomplete collapse, bridging the gap between discrete-time models and continuous-time intrinsic bubbles.

Furthermore, empirical literature on bubble detection—most notably the right-tailed unit root tests developed by \cite{phillips2011explosive} and \cite{phillips2015testing}—has robustly confirmed that many economic and financial time series periodically exhibit locally explosive behavior (bubble formation) followed by crashes. 
Motivated by these empirical facts, nonlinear time-series models have gained attention for directly capturing such dynamics. 
Recently, \cite{yang2024bubble} proposed a novel and simple time-series model known as the stochastic nonlinear autoregressive (SNAR) model. 
Their work demonstrates that, despite operating within a strictly stationary process framework, this model can accurately portray bubble dynamics characterized by periodic local explosions and subsequent collapses.

These preceding studies have made significant contributions to the mathematical and statistical description of bubble periodicity and collapse mechanisms. 
However, there remains room for exploration in developing models that integrate the micro-level dynamics of actual market trading behaviors into an extremely simple and endogenous mechanism. 
In the field of econophysics, approaches such as \cite{bouchaud1998langevin} have demonstrated the utility of Langevin-type equations to describe stock market fluctuations and crashes driven by collective investor behavior. 
Yet, synthesizing these micro-behavioral insights—specifically,``investor herd psychology (trend following)" and ``rapid panic selling triggered by fear during market overheating"—with the discrete-time periodic bubble literature remains a challenge.

Therefore, this paper proposes a new discrete-time simulation model for financial bubbles, eliminating complex assumptions to prioritize simplicity as its defining feature. 
The primary innovation of this model is that the balance between investors' buy orders and sell orders is determined by a cubic function of momentum $M_t$, which represents the accumulation of past price fluctuations, as employed in \cite{lin2019simple}.
When the momentum is within a moderate range (indicative of trend-following behavior), positive feedback drives bubble formation. However, once the momentum exceeds a certain threshold and the market overheats, the properties of the cubic function endogenously generate strong selling pressure, triggering a dramatic bubble collapse.

Furthermore, inspired by the Hawkes process \cite{hawkes1971spectra}, the model incorporates a mechanism wherein the frequency of trading itself increases in response to momentum. This feature allows the model to naturally replicate the ``market frenzy," during which volatility and liquidity surge as the bubble forms.

The remainder of this paper is structured as follows. Section 2 presents the formalization of the proposed simulation model, explaining the price-updating mechanism driven by the momentum-dependent cubic function and Section 3 provides the simulation result of this basic model. Section 4 analyzes the impact of various model parameters—such as the momentum threshold $b$, the memory decay rate $r$, and the baseline trading probability $\Lambda$—on the scale and periodicity of the bubbles. Finally, Section 5 provides the concluding remarks.

\section{Model}

In this section, we present a simple discrete-time simulation model to describe the endogenous formation and collapse of financial bubbles. The model consists of two main mechanisms driven by market momentum: the acceleration of trading frequency (inspired by the Hawkes process) and the nonlinear transition of price directions using a cubic function (inspired by the potential in the Langevin equation).

Let $P_t$ be the price of an asset at time $t$ for $t = 1, 2, \dots, T$. We define the market momentum $M_t$ as the exponentially weighted moving average of past log-returns:
\begin{equation}
    M_t = \sum_{s=1}^{t-1} e^{-r(t-s)}(\log P_s - \log P_{s-1}),
\end{equation}
where $r > 0$ determines the decay rate of past price changes. A smaller $r$ implies that investors have a longer memory of the historical trend.

\subsection{Trading Frequency Driven by Momentum}
During a bubble, market enthusiasm typically leads to a surge in trading volume. To capture this phenomenon, we define the intensity of trading occurrence, $\lambda_t$, as a linear function of the momentum $M_t$:
\begin{equation}
    \lambda_t = \Lambda + k M_t,
\end{equation}
where $\Lambda$ is the baseline parameter for the minimum trading probability, and $k > 0$ determines the sensitivity of the trading intensity to the momentum. 

Let $\Delta N_t \in \{0, 1\}$ be an indicator variable taking the value 1 if a trade occurs at time $t$, and 0 otherwise. We assume that $\Delta N_t$ follows a Bernoulli distribution:
\begin{equation}
    \Delta N_t \sim Be(\Phi(\lambda_t)),
\end{equation}
where $\Phi(\cdot)$ denotes the cumulative distribution function (CDF) of the standard normal distribution. Consequently, the cumulative number of trades up to time $t$ is given by $N_t = N_{t-1} + \Delta N_t$. This formulation ensures that as the momentum builds up, the probability of trading increases, endogenously generating market liquidity.

\subsection{Nonlinear Price Direction and Bubble Collapse}
The core innovation of our model lies in the determination of the price direction. We introduce a state variable $x_t$ that governs the balance between buying and selling pressures. The evolution of $x_t$ is defined by a cubic polynomial of the momentum:
\begin{equation}
    x_t = x_{t-1} + h(M_t - a)(M_t - b)(M_t - c),
\end{equation}
where $h > 0$ is a scaling parameter, and $a, b, c$ ($a < b < c$) are the roots of the cubic function. 

This cubic specification models the nonlinear herd behavior of investors, analogous to the movement within a potential landscape in the Langevin equation (see, e.g., \cite{bouchaud1998langevin}). When $M_t$ is slightly positive ($0 < M_t < b$), the cubic term is likely to be positive, pushing $x_t$ upward. This represents a trend-following behavior where positive momentum encourages further buying. However, when the momentum exceeds the critical threshold $b$, the term $(M_t - b)$ becomes positive, causing the entire cubic function to turn negative. This mechanism elegantly captures the sudden phase transition in market sentiment: an overheated market triggers panic selling, leading to a sharp crash.

The actual direction of the price change, $Z_t \in \{0, 1\}$ (where 1 indicates a price increase and 0 indicates a decrease), is generated from a Bernoulli distribution dependent on $x_t$:
\begin{equation}
    Z_t \sim Be(\Phi(x_t)).
\end{equation}

\subsection{Price Update Rule}
Finally, the log-price is updated only when a trade occurs ($\Delta N_t = 1$). The price changes by a constant tick size $d$:
\begin{equation}
    \log P_t = \log P_{t-1} + d (2Z_t - 1) (N_t - N_{t-1}).
\end{equation}
Since $Z_t \in \{0, 1\}$, the term $(2Z_t - 1)$ outputs either $+1$ or $-1$, ensuring a simple random walk-like structure with an endogenously varying drift.

\section{Simulation}

\subsection{Initial Values and Parameter Settings}
For the baseline simulation, we set the total number of periods to $T = 5000$. The initial states are set as $\log P_0 = \log P_1 = 0$ and $x_0 = x_1 = 0$. 

The baseline parameters are configured as follows: 
the tick size $d = 0.01$, the memory decay rate $r = 0.001$, the baseline probability $\Lambda = -2$, and the momentum sensitivities $k = 10$ and $h = 0.2$. 

For the cubic function roots, we set $a = -1$, $c = 1$ (which are sufficiently large in absolute terms), and the critical crash threshold $b = 0.02$, implying that selling pressure dominates when the momentum reaches twice the base tick size.

\subsection{Simulation result}

Figure 1 illustrates the simulation results under the baseline parameter settings. The four panels display, from top to bottom, the evolution of the log-price $\log P_t$, the market momentum $M_t$, the trading intensity $\lambda_t$, and the state variable for price direction $x_t$.

As observed in the top panel, an upward trend gradually forms, leading to a ``bubble" characterized by a sharp surge in the log-price, which is then followed by an instantaneous and dramatic collapse (crash). This nonlinear dynamics is driven by the internal mechanisms of the proposed model, which can be intuitively understood through the remaining three panels.

First, the second and third panels demonstrate the ``overheating of trading frequency." As the price rises, positive momentum $M_t$ accumulates. Consequently, the trading intensity $\lambda_t$ linearly increases, elevating the probability of trade occurrences. This mechanism successfully captures the ``market frenzy" phase of a bubble, where market participants rush in, causing an explosive increase in market liquidity and volatility.

Second, the relationship between the second and fourth panels clearly reveals the ``endogenous phase transition and panic selling" induced by the cubic function. During the period of moderate momentum ($0 < M_t < b$), the state variable $x_t$ continues to rise, reflecting strong trend-following (herd) behavior among investors. However, the exact moment the market excessively overheats and the momentum $M_t$ exceeds the critical threshold $b$ (indicated by the red dashed line in the second panel), the property of the cubic function causes $x_t$ to plunge deeply into the negative region. This indicates that overwhelming selling pressure instantly dominates the market, triggering a sudden bubble crash.

Through this visualization of the internal state variables, it is clearly confirmed how the model endogenously generates complex and nonlinear market dynamics: the overheating of the market driven by positive feedback, and the subsequent sharp crash triggered mechanically by exceeding the momentum threshold.

\begin{figure}[htbp]
  \centering
  \includegraphics[width=\linewidth]{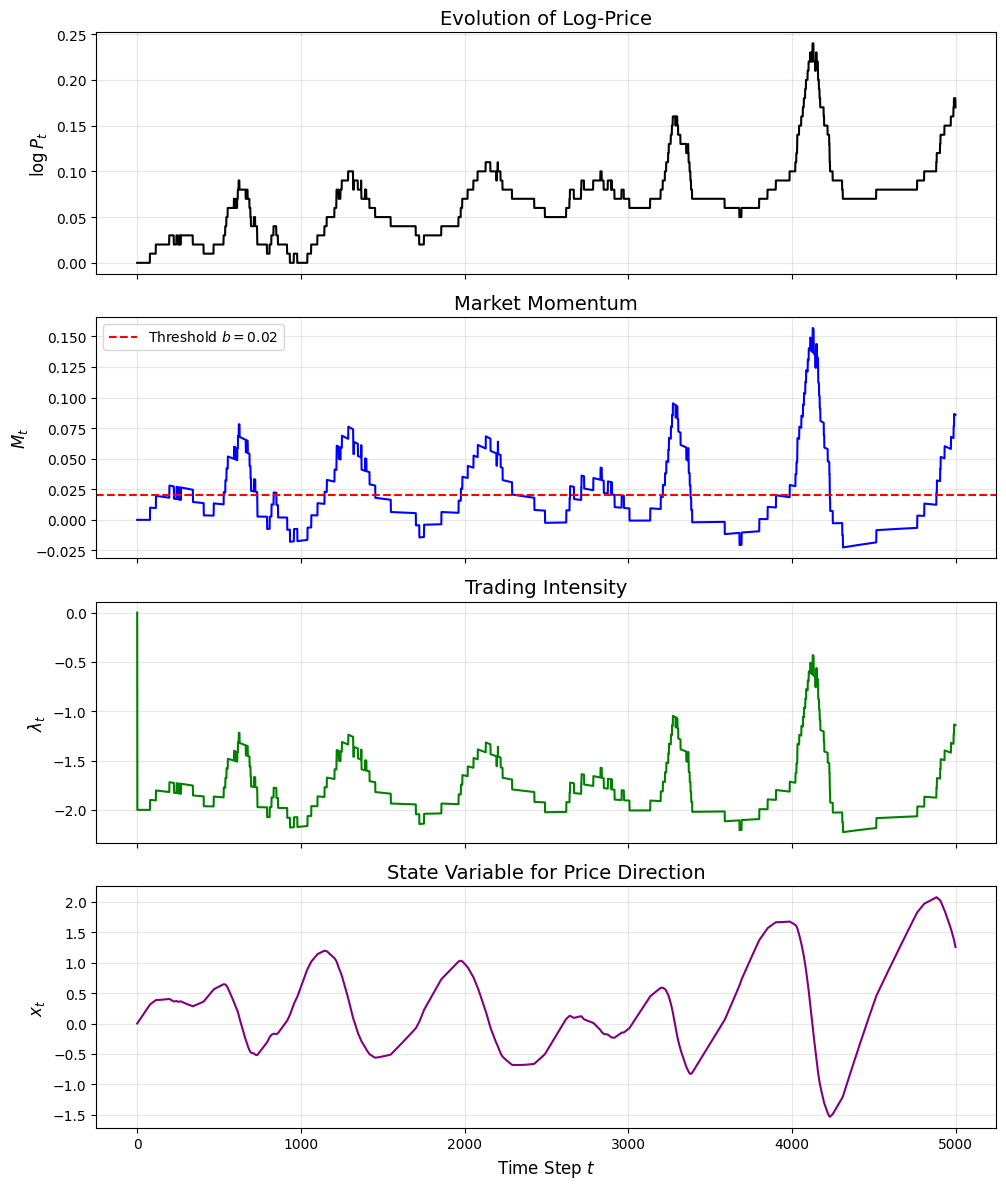}
  \caption{Simulation results of bubble formation and collapse under baseline parameters. The four panels display, from top to bottom: the log-price $\log P_t$, the market momentum $M_t$ with the crash threshold $b$ (red dashed line), the trading intensity $\lambda_t$, and the state variable for price direction $x_t$.}
  \label{fig:sim1}
\end{figure}

\subsection{Parameter Sensitivity Analysis}

Figure 2 presents the sensitivity of the log-price dynamics to three key parameters: the crash threshold $b$, the memory decay rate $r$, and the baseline trading probability $\Lambda$.

\subsubsection{Impact of Crash Threshold $b$}
The top panel of Figure 2 illustrates how the crash threshold $b$ determines the ``size of the game of chicken." When $b$ is extremely low ($b=0.0001$), the market is hypersensitive to any upward movement. The momentum $M_t$ hits the threshold almost immediately after a trend begins, triggering the cubic function's selling pressure prematurely. As a result, only micro-bubbles with negligible price increases are observed. In contrast, as $b$ increases to $0.01$ and $0.02$, the market develops a higher tolerance for overheating. This allows the momentum to accumulate significantly, enabling the log-price to reach much higher peaks before the endogenous transition to a crash occurs. This result highlights $b$ as a parameter representing the market's risk appetite or the structural limit of investor confidence.

\subsubsection{Impact of Memory Decay Rate $r$}
The middle panel shows a non-linear ``Goldilocks effect" regarding the memory decay rate $r$. A high decay rate ($r=0.005$) prevents bubble formation because the momentum ``leaks" too fast to ignite the self-exciting trading loop. Conversely, an extremely low decay rate ($r=0.0005$) leads to premature crashes because the market remembers all past gains perfectly, causing $M_t$ to hit the threshold $b$ too efficiently. The most explosive and sharp bubbles occur at an intermediate level ($r=0.001$), where moderate forgetting acts as a ``pressure release valve," delaying the crash and allowing the price to soar to its maximum potential.

\subsubsection{Impact of Baseline Trading Probability $\Lambda$}
The bottom panel displays the effect of the baseline trading probability $\Lambda$ on the frequency of bubble occurrences. A higher $\Lambda$ (e.g., $-1.5$) represents a more active market with higher baseline liquidity. In such an environment, small price fluctuations are more likely to occur, which frequently serve as ``seeds" for the positive feedback loop of momentum, leading to more frequent bubble cycles. On the other hand, a lower $\Lambda$ (e.g., $-2.5$) describes a ``cold" market where transactions are sparse, making it more difficult for a sustained trend to ignite, thereby increasing the intervals between bubble events.

\begin{figure}[htbp]
  \centering
  \includegraphics[width=0.95\linewidth]{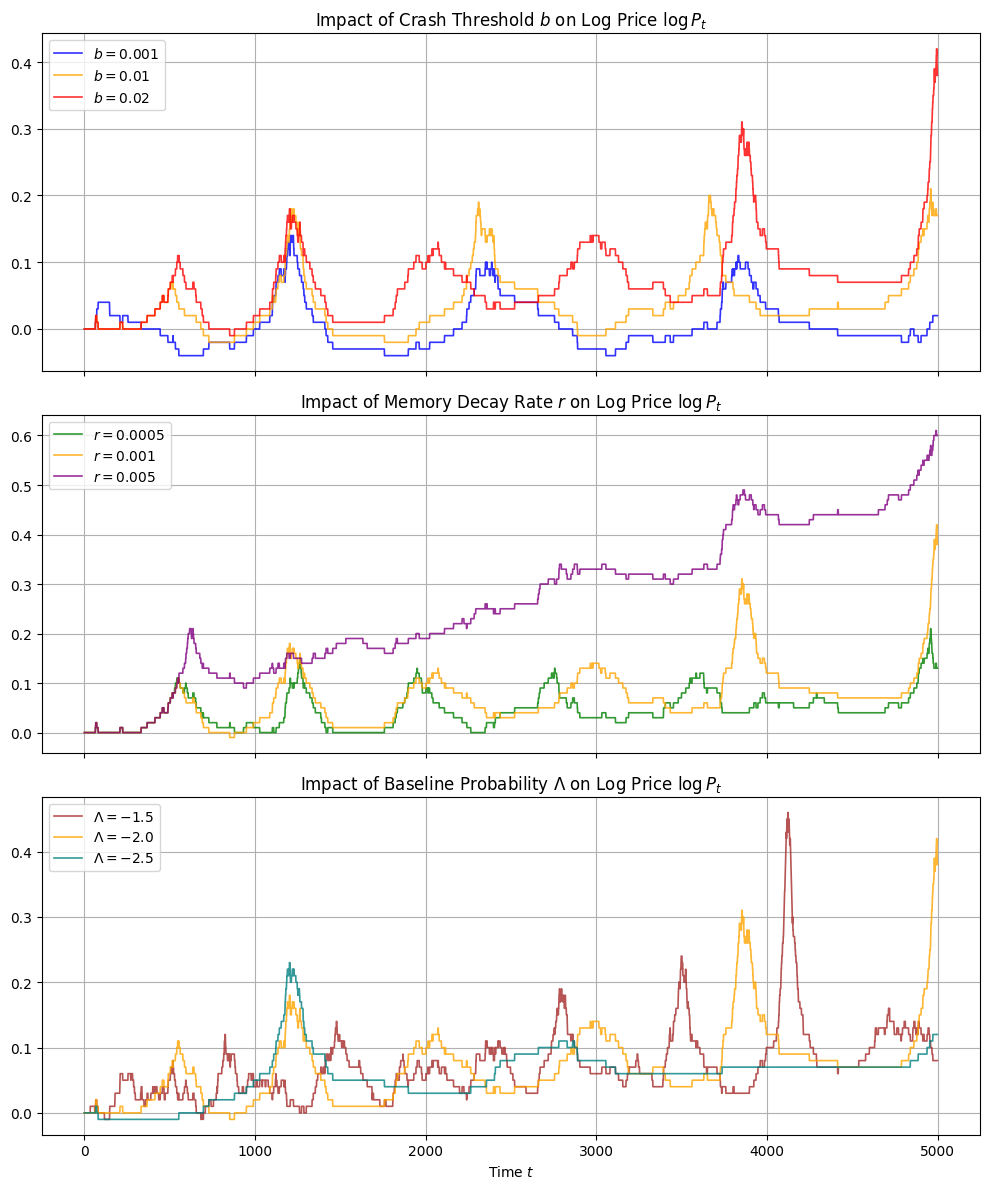}
  \caption{Impact of key model parameters on bubble dynamics (evolution of log-price $\log P_t$). Each panel illustrates the simulation results when only a single parameter is varied from the baseline setting ($d=0.01, a=-1, c=1, k=10, h=0.2$). The top panel shows the effect of the momentum threshold for crash $b$, the middle panel shows the effect of the memory decay rate $r$, and the bottom panel shows the effect of the baseline trading probability $\Lambda$.}
  \label{fig:sim2}
\end{figure}

\section{Conclusion}

This paper has proposed a novel and parsimonious discrete-time simulation model to describe the endogenous formation and collapse of financial bubbles. 
By integrating investors' micro-level behaviors—specifically, trend-following herd mentality and panic selling during market overheating—into a simple mathematical framework using a cubic function and momentum-dependent trading frequency, we have successfully replicated complex, nonlinear market dynamics that have historically been difficult to capture in a unified, simple model.
The key findings and contributions of this study are summarized as follows. 
First, the model endogenously generates periodic bubbles and crashes without requiring exogenous shocks or complex regime-switching assumptions. 
The cubic function of momentum acts as a potential landscape that allows for a natural phase transition: while moderate momentum encourages further buying (the bubble phase), exceeding a critical threshold $b$ triggers an immediate and overwhelming shift to selling pressure (the crash phase). 
Second, by incorporating a Hawkes-inspired mechanism where trading frequency accelerates with momentum, the model successfully reproduces the ``market frenzy" observed in real-world bubbles, characterized by a simultaneous surge in liquidity and price. 
Third, the parameter impact analysis demonstrated that the scale and periodicity of bubbles are highly sensitive to investor memory ($r$), tolerance for overheating ($b$), and baseline market liquidity ($\Lambda$), providing intuitive economic interpretations for various market regimes.

Despite its simplicity, the proposed model offers a robust platform for understanding the qualitative nature of financial instability. 
However, several avenues for future research remain. 
First, establishing a rigorous parameter estimation method is essential. 
While this study focused on qualitative simulations, developing quantitative estimation techniques—such as maximum likelihood estimation or the method of simulated moments—using high-frequency empirical market data will be a crucial next step. 
Second, extending the trading frequency mechanism is a highly promising direction. 
While the current model incorporates a linear feedback mechanism inspired by the standard Hawkes process, replacing it with a Quadratic Hawkes process (see, e.g., \cite{blanc2017quadratic,fosset2022non}) could allow the model to capture the more explosive, non-linear self-exciting nature of market volatility typically observed during extreme bubble formations. 
Finally, extending the model into a multi-agent framework to explore the interactions between different types of investors (e.g., fundamentalists and chartists) could provide further insights into the structural causes of market fragility.

In conclusion, the strength of this model lies in its ability to generate rich, nonlinear phenomena from minimal components. 
By focusing on the interplay between momentum and trading intensity, this research contributes a new perspective to the ongoing effort to model and eventually anticipate the volatile cycles of financial markets.

\bibliography{ref} 
\bibliographystyle{apalike} 

\end{document}